\documentclass[twocolumn]{aastex631}

\accepted{in ApJ}

\shorttitle{Filaments and Hub Formation in G083.097$+$03.270}

\shortauthors{Panja et al.}

\graphicspath{{./}{figures/}}

\begin{document}

\title{Merging Filaments and Hub Formation in the G083.097$+$03.270 Molecular Complex}

\correspondingauthor{Alik Panja}
\email{alik.panja@gmail.com}

\author[0000-0002-4719-3706]{Alik Panja}
\affiliation{S. N. Bose National Centre for Basic Sciences, Sector-III, Salt Lake, Kolkata 700106, India}

\author[0000-0001-6725-0483]{Lokesh K. Dewangan}
\affiliation{Astronomy \& Astrophysics Division, Physical Research Laboratory, Navrangpura, Ahmedabad 380009, India}

\author[0000-0003-0295-6586]{Tapas Baug}
\affiliation{S. N. Bose National Centre for Basic Sciences, Sector-III, Salt Lake, Kolkata 700106, India}

\author[0000-0003-0262-272X]{Wen Ping Chen}
\affiliation{Institute of Astronomy, National Central University, 300 Zhongda Road, Zhongli, Taoyuan 32001, Taiwan}
\affiliation{Department of Physics, National Central University, 300 Zhongda Road, Zhongli, Taoyuan 32001, Taiwan}

\author[0000-0002-3904-1622]{Yan Sun}
\affiliation{Purple Mountain Observatory, Chinese Academy of Sciences, 10 Yuanhua Road, Nanjing 210033, China}

\author[0000-0001-5508-6575]{Tirthendu Sinha}
\affiliation{S. N. Bose National Centre for Basic Sciences, Sector-III, Salt Lake, Kolkata 700106, India}

\author[0000-0003-1457-0541]{Soumen Mondal}
\affiliation{S. N. Bose National Centre for Basic Sciences, Sector-III, Salt Lake, Kolkata 700106, India}


\begin{abstract}

We uncover a hub-filament system associated with massive star formation in the G083.097$+$03.270.  Diagnosed with simultaneous $^{12}$CO, $^{13}$CO, and C$^{18}$O line observations, the region is found to host two distinct and elongated filaments having separate velocity components, interacting spatially and kinematically, that appear to have seeded the formation of a dense hub at the intersection. 
A large velocity spread at the hub in addition to clear bridging feature connecting the filaments in velocity are indicating merging of filaments.  
Along the filaments axis, the velocity gradient reveals a global gas motion with an increasing velocity dispersion inward to the hub signifying turbulence.  Altogether, the clustering of Class~I sources, a high excitation temperature, a high column density, and presence of a massive outflow at the central hub suggest enhanced star formation.  
We propose that merging of large-scale filaments and velocity gradients along filaments are the driving factors in the mass accumulation process at the hub that have sequentially led to the massive star formation.  
With two giant filaments merging to coincide with a hub therein with ongoing star formation, this site serves as a benchmark for the `filaments to clusters' star-forming paradigm.  

\end{abstract}


\section{Introduction} \label{sec:intro}

The filamentary structures are ubiquitous in the interstellar medium \citep{andre10, mol10}.  
Enlightened by the Herschel Space Mission \citep{pil10} and the succeeding studies have revealed that majority of the dense ($A_{V}>7$~mag) molecular gas involved in star formation is dispersed in filamentary formations \citep{andre10, arz19, hac22, pin22}.  
Those with higher column densities ($N_{\mathrm{H_{2}}} \gtrsim 10^{22}$~cm$^{-2}$) are preferred sites to initiate star formation, when the densest filaments due to gravitational instability are subject to fragmentation into prestellar cores.  
Filaments exhibit a wide range of physical scales and kinematic properties \citep[see the recent reviews by][]{hac22, pin22}, leading subsequently to different dynamical evolution.  

Filaments are often found networked in complexity, where multiple units with sizes several parsecs radially merging into a central parsec-scale clump, referred to as hubs, which have low aspect ratios and high-column densities \citep{mye09}.  These hub-filament systems (HFSs) are considered as the potential progenitors of massive young stellar associations, in which luminous ($>10^{4}$~L$_{\sun}$) massive stars are formed \citep{kum20}.   

A comprehensive evolutionary sequence of the HFS has been forwarded by \citet{kum20}, in which the formation scenario of massive stars from molecular gas are categorized in snapshots of four consecutive stages.  Briefly, in Stage~I, dense filaments move towards each other and set up the initial conditions for HFS formation. The formation of HFS can be initiated by a variety of processes, manifest as flow-driven filaments, energetic stellar wind bubbles, expanding ionization fronts, supernova shocks, etc. \citep{kum20}.  In Stage~II, the approaching filaments merge and form the hub with a small twist at the overlapping zone, thereby flattening the hub.  In Stage~III, the density in the hub amplifies due to the initial shock followed by self-gravity, and hence drives longitudinal flows toward the hub leading to formation of massive stars.  Finally in Stage~IV, the radiation pressure and ionization feedback from the massive stars shape the remnant filaments as pillars, leaving a mass-segregated embedded cluster at the hub.  It has been rapidly recognized that the HFS may play an important, if not dominant, role in massive star and cluster formation \citep{schneider12, baug18, montillaud19b, kum22}.  

However, till date, direct observational evidence validating the causality of HFS with cluster formation has only emerged in a handful of studies in the literature.  In addition to the results presented in \citet{kum20, kum22}, only a few cases have so far been reported, e.g., in the Mon\,OB1 star-forming region, \citet{montillaud19b} demonstrated a pair of filaments approaching each other (Stage~I), and recently as another example, in the massive protocluster G31.41$+$0.31, \citet{beltran22} identified cloud structures that represent a transition phase between Stage~III and Stage~IV.  
In this work, we present an observational scenario of two giant molecular filaments crossing each other and at the intersection forming a hub along with noticeable presence of a group of protostars, resembling Stage~II/III of the `filaments to clusters' model \citep{kum20}.  

The target of our study is G083.097$+$03.270 (hereafter G08, $\ell = 83\fdg0970$; $b = +03\fdg2700$), which is a part of the large scale area previously studied by \citet{pan22} and therein referred to as Clump~B.  This complex, located in the Orion arm near the Cyg~X system, was found by these authors associated with intense thermal dust emission (AKARI: 160~$\mu$m; Planck: 353~GHz), strong ionized gas emission (NVSS: 1.4~GHz), and dense molecular cloud (PMO: $^{12}$CO, $^{13}$CO, C$^{18}$O).  Moreover, the large scale ($\sim 2\fdg0$) molecular gas is connected spatially and kinematically with a nearby nebulous \ion{H}{2} complex Sh2-112, which has a distance of $\sim 2.1$~kpc.  \citet{and15} studied the radio recombination lines and referred to this \ion{H}{2} region as G083.097$+$03.270, which is related with the molecular cloud 083.1$+$03.3 \citep{dob94}.  A massive young stellar object \citep[VLA\,G083.0934$+$03.2720;][]{urq09} detected from the Very Large Array (VLA) radio continuum observations coincides with the central region, where a significant velocity dispersion has been reported.  \citet{pan22} also presented ionized and molecular gas morphology and parameters, and inferred the central region to likely host a group of ionizing stars.  

An inspection of the molecular gas kinematics in G08 reveals spatially connected filaments having separate velocity components.  This site therefore could be a potential HFS candidate that encompasses a dynamic range of substantial activities, from gas motions within filaments to young massive stellar cluster formation.  Here we present a thorough investigation on the interaction of filaments and their dynamical evolution to clusters based on high spectral resolution (0.16--0.17~km~s$^{-1}$) observational data.  We focus on the analysis of a relatively compact complex ($0\fdg9 \times 0\fdg9$) with a narrow velocity range ([$-6.0$, $+1.0$]~km~s$^{-1}$).  

The paper is organized as follows.  In Section~\ref{sec:data} we present the observational data used in this work, followed by Section~\ref{sec:results} of the spatial and kinematic morphology, and hence the possible interplay, of the filaments.  In Section~\ref{sec:discuss} we discuss the plausible formation history of the hub influenced by merging of filaments and their observable outcomes.  Finally, Section~\ref{sec:summary} contains a summary of the main results of this work.

\section{Data} \label{sec:data}

We have used the $^{12}$CO, $^{13}$CO, and C$^{18}$O ($J$=1--0) molecular line data obtained from the Milky Way Imaging Scroll Painting \citep[MWISP;][]{su19} survey carried out by the 13.7~m diameter millimetre-wavelength telescope of the Purple Mountain Observatory (PMO), China.  For simultaneous observations of the three CO isotopologues, a multibeam sideband-separating Superconducting Spectroscopic Array Receiver system with an instantaneous bandwidth of 1~GHz is employed.  The typical system temperatures are $\sim 250$~K (rms noise $\sim 0.5$~K) for $^{12}$CO at the upper sideband, and $\sim 140$~K (rms noise $\sim 0.3$~K) for $^{13}$CO and C$^{18}$O at the lower sideband.  
Finally the raw data are resampled and mosaicked into FITS cubes for a spatial resolution of $30\arcsec$ and a velocity resolution of 0.16--0.17~km~s$^{-1}$.  We have reprocessed a part of the data already presented in \citet{pan22} in the velocity range [$-6.0$, $+1.0$]~km~s$^{-1}$.  The observational techniques and reduction methodology are further detailed there.

\section{Results} \label{sec:results}

Here we have segregated the filament components based on spatial and spectral properties, using the observed CO isotopologues.  We then have diagnosed the pattern of the gas flow within filaments and the nature of possible interaction between them.

\subsection{Hub-Filament Morphology in G08} \label{subsec:morphology}

The molecular gas spatial and kinematic structures are diagnosed using the CO isotopologues.  The $^{12}$CO emission is optically thick, therefore suitable to trace the spatial extents of the diffuse extended gas (density $\sim 10^2$~cm$^{-3}$).  On the other hand, $^{13}$CO and C$^{18}$O trace comparatively dense ($\sim 10^{3}$--$10^{4}$~cm$^{-3}$) structures, except that C$^{18}$O is optically even thinner than $^{13}$CO hence reveals denser regions than $^{13}$CO does.  Thus for this study, we have relied more on the $^{13}$CO data to decipher the relatively dense hub and analogous filament structures.

\subsubsection{Integrated Intensity}
 \label{subsubsec:intensity}

The $^{13}$CO integrated intensity map for the velocity range [$-6.0$, $+1.0$]~km~s$^{-1}$ toward G08 is shown in Figure~\ref{fig:g_moment012}(a), which clearly reveals two elongated filaments (hereafter Fi-N and Fi-S), aligned in a tilted Y-shaped distribution along the Galactic east-west direction.  We recognize these structures as `giant filaments' categorized in the {\textit {filament families}} of \citet{hac22}.  The two filaments interlace spatially, forming a hub, wherein we observe a notable enhancement in intensity.  Toward the western part of Fi-N, we also see intensification at certain regions.

\subsubsection{Velocity}
 \label{subsubsec:velocity}

The velocity distribution ($^{13}$CO) of the molecular gas, shown in Figure~\ref{fig:g_moment012}(b), reveals two separate components.  The median velocities of Fi-N and Fi-S are found to be $-3.98$~km~s$^{-1}$ and $-0.83$~km~s$^{-1}$ considering their entire coverage area, estimated from the intensity-weighted respective $^{13}$CO line width.  A continuous velocity gradient is detected along Fi-N, suggesting inflow motions most likely toward the center of gravity, with bluer velocity along Galactic east ($\sim -4.31$~km~s$^{-1}$) to relatively redder in west ($\sim -2.41$~km~s$^{-1}$).  However we caution that, the velocity ranges for the filaments mentioned above are to considered as representative velocities and not as definite values, particularly for Fi-N, which displays a significant shift in peak velocity from east to west.

\subsubsection{Velocity Dispersion}
 \label{subsubsec:dispersion}

The velocity dispersion map ($^{13}$CO) is presented in Figure~\ref{fig:g_moment012}(c), that hints on a very turbulent gas motion around the hub.  We mapped the median velocity dispersion around the junction as $\sim 3.0$~km~s$^{-1}$, a dispersion that could likely be caused by supersonic gas flows, which would inevitably create strong shock compression at the junction of filaments.  
From Figure~\ref{fig:g_moment012}(a)-(c), it is also distinctly evident that Fi-S has actually an extended arm to the north-eastern, whereas at the intersection of two filaments there is an X-shaped structure, also visible in Figure~9 of \citet{pan22}.

\subsubsection{H$_{2}$ Column Density}
 \label{subsubsec:column}

\begin{figure*}
\centering
        \includegraphics[width=\textwidth]{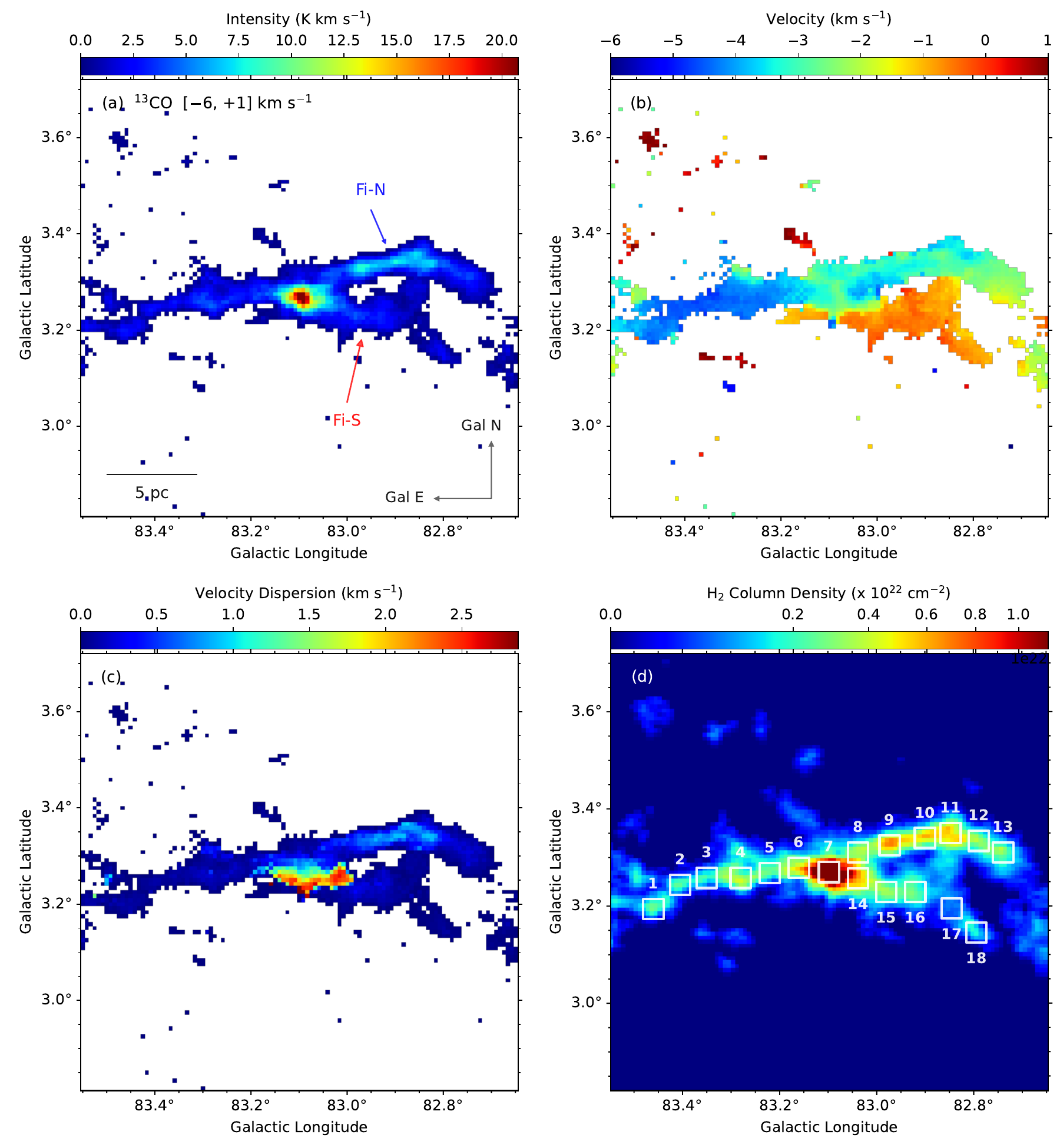}
  \caption{(a) The intensity weighted moment-0 map for $^{13}$CO ($J$=1--0), integrated in the velocity range [$-6.0$, $+1.0$]~km~s$^{-1}$ toward G083.097$+$03.270 (G08).  Two filaments, Fi-N and Fi-S, are revealed and are found to be intersecting, forming a hub, i.e., with an enhanced gas density at the center. (b) The $^{13}$CO moment-1 map shows the filaments to have two different velocity structures.  (c) In the $^{13}$CO moment-2 map, the large velocity dispersion is traced mostly in and around the junction of filaments.  (d) The H$_{2}$ column density derived from $^{13}$CO for the molecular gas with an overplot of square regions each of size $2\farcm5$ and numbered consequently, selected to study the variation of molecular properties along the filaments axis.  Panels  (a) to (c) are in linear scaling, whereas (d) is in square root scaling. }
  \label{fig:g_moment012}
\end{figure*}

The H$_{2}$ column density map integrated for the mentioned velocity range and derived from $^{13}$CO is depicted in Figure~\ref{fig:g_moment012}(d). The highest column density occurs at the hub, reaching $\sim 4.8 \times 10^{22}$~cm$^{-2}$, which surpasses the typical threshold of $\sim 1.0 \times 10^{22}$~cm$^{-2}$ for massive star formation \citep{kru08}.  Additionally in this map, we have selected a few square regions, each of size $2\farcm5$ (5 pixels), along the filament axis for the average spectra to investigate possible velocity entanglements, as will be discussed in the following section.  
The specific locations of the box are chosen along the filaments axis at those zones where we found relatively higher intensity ($^{13}$CO) visually.

\subsubsection{Distance}
 \label{subsubsec:distance}

The distance is an essential parameter to verify if the filaments are parts of the same physical system.  In order to derive the distance of G08, we have utilized the BeSSeL \citep{rei19} Survey, that computes the trigonometric parallaxes of massive star forming regions by taking into account of the spiral structure and kinematics of the Milky Way.  Considering a median radial velocity of $-3.98$~km~s$^{-1}$ for Fi-N and $-0.83$~km~s$^{-1}$ for Fi-S, the model produced kinematic distances of $\sim 1.53 \pm 0.12$~kpc for Fi-N, and $\sim 1.51 \pm 0.13$~kpc for Fi-S, in the Local arm, placing the filaments at the same plane.  
To validate this distance measurement with another approach, we also made use of the Galactic three-dimensional dust distribution  \citep{gre19}, based on Gaia parallaxes, and stellar photometry from Pan-STARRS~1 and 2MASS.  But unlikely, this model produced two ranges of distances ($\sim 1.33$~kpc and $\sim 2.52$~kpc) for the region, deviating from those of the kinematic values.  Therefore we have relied on the Galactic rotation model and used an average distance of $\sim 1.52 \pm 0.12$~kpc for our subsequent analysis.  We note that the kinematic distance measurements for nearby local clouds could be associated with a certain level of uncertainty.

\subsection{Kinematics of Molecular Gas Along the Filaments} \label{subsec:kinematics}

\subsubsection{Average Spectra}
 \label{subsubsec:spectra}

\begin{figure*}
\centering
        \includegraphics[width=\textwidth]{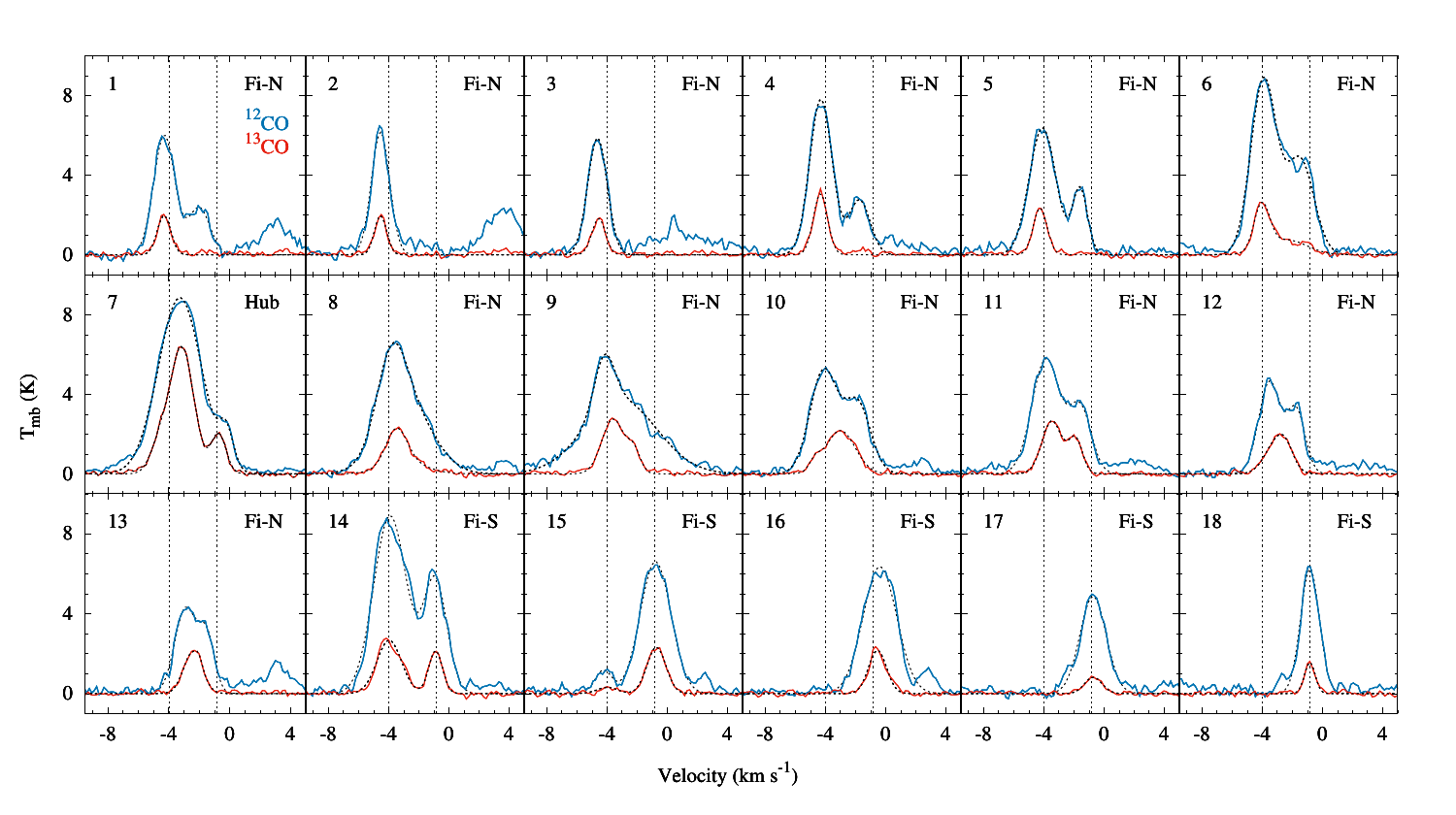}
  \caption{Average spectra (blue: $^{12}$CO; red: $^{13}$CO) fitted with Gaussian (black dashed curves) for the square regions shown in Figure~\ref{fig:g_moment012}(d), each labelled with an ID number.  The vertical dashed lines mark velocities of $-3.98$~km~s$^{-1}$ and $-0.83$~km~s$^{-1}$, respectively.  For ID 7 and $^{12}$CO, the $T_{\mathrm{mb}}$~(K) is scaled down by a factor of 2.  }
  \label{fig:g_spectra}
\end{figure*}

The average spectra along Fi-N and Fi-S for the square regions (mentioned in Section~\ref{subsubsec:column} and marked in Figure~\ref{fig:g_moment012}(d)) for both $^{12}$CO and $^{13}$CO are shown in Figure~\ref{fig:g_spectra}.  For region IDs 1 to 13 correspond to Fi-N (7 being the hub) and 14--18 to Fi-S.  For ID up to 3, a single peak related with the $\sim -3.98$~km~s$^{-1}$ cloud dominates in both $^{12}$CO and $^{13}$CO.  But from ID 4 to 7 both the peaks ($\sim -3.98$~km~s$^{-1}$ and $\sim -0.83$~km~s$^{-1}$) are visible in $^{12}$CO, while both are seen near the hub (ID 6 and 7) in $^{13}$CO.  For IDs 8 and 9, complex structures dominated by single peaks with large spread are prominent.  Again from ID 10 to 13, both the peaks can be seen in $^{12}$CO, but only for ID 11 in $^{13}$CO.  For ID 14 and 15, both near the hub, the peaks are present, but for IDs from 16 to 18, the $-0.83$~km~s$^{-1}$ peak prevails.  To summarize, for  regions adjacent to the hub (ID 6, 7, 14), eminent presence of both the velocity components in both $^{12}$CO and $^{13}$CO stipulates conjunction of molecular gas from two different velocity filaments.

\subsubsection{Position-Velocity Diagrams}
 \label{subsubsec:pvd}

\begin{figure*}
\centering
        \includegraphics[width=1.0\textwidth]{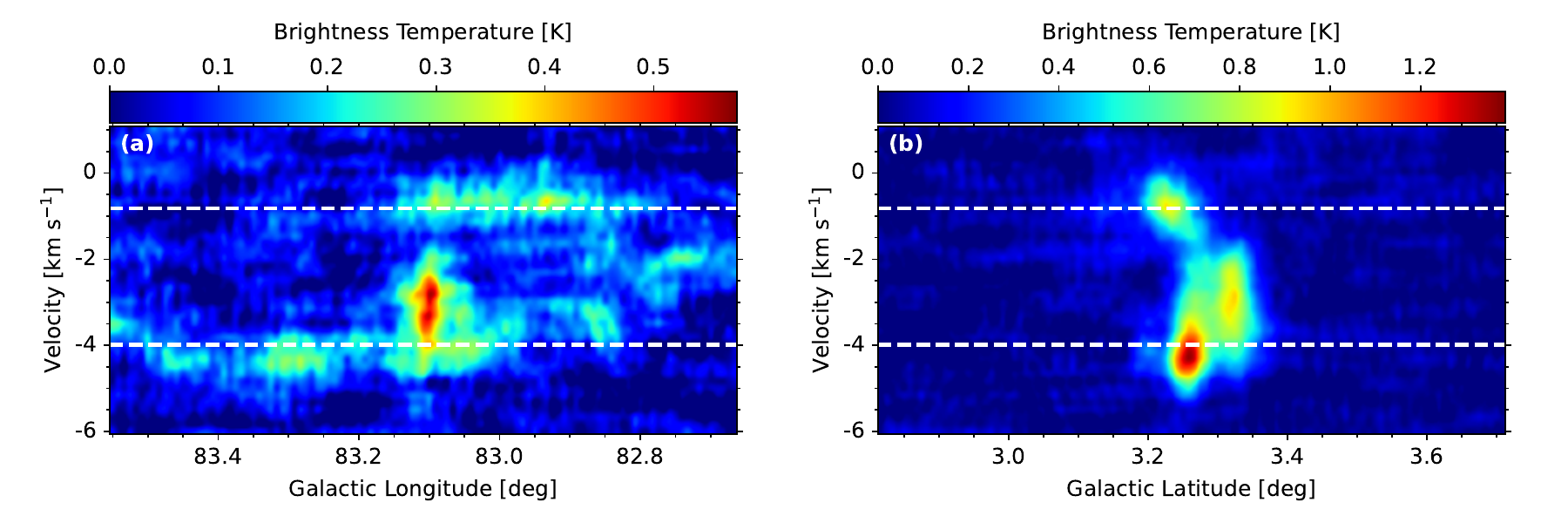}
  \caption{The $^{13}$CO position-velocity diagrams toward G08.  (a) Longitude-velocity distribution is integrated for the entire latitude range [$2\fdg82$, $3\fdg72$] and (b) latitude-velocity distribution is integrated for the entire longitude range [$83\fdg547$, $82\fdg647$].  The white dashed lines mark the  velocity of $-3.98$~km~s$^{-1}$ and $-0.83$~km~s$^{-1}$, respectively.  }
  \label{fig:g_lbv}
\end{figure*}

Another observational signature, if the filaments are indeed connected in velocity, is to probe the position-velocity diagram.  As shown in the $^{13}$CO longitude-velocity map, in Figure~\ref{fig:g_lbv}(a), the Fi-N and Fi-S comprise two distinct velocity components ($\sim -3.98$~km~s$^{-1}$ and $\sim -0.83$~km~s$^{-1}$).  The molecular gas in the two filaments are most prominently linked by the bridge feature (intermediate velocity) at the hub.  Also we witness a few additional locations of inter-filamentary interaction between the western part of Fi-N (segmented as Fi-NW, hereafter) and Fi-S, therefore shifting the velocity in Fi-NW to an intermediate range.  From now on we use the notation that Fi-N consists of two segments, one is eastern part (hereafter Fi-NE) and another is western (Fi-NW), showing slightly different kinematics.  In the $^{13}$CO latitude-velocity map in Figure~\ref{fig:g_lbv}(b), Fi-N and Fi-S are clearly traced with separate velocities connected by the bridge feature.  In this map, though there is a significant gas overlap between Fi-NE and Fi-NW kinematically, the velocity peaks differ marginally, probably due to the mutual interaction between Fi-NW and Fi-S.

\subsubsection{Channel Maps and Interaction}
 \label{subsubsec:channel}

The Fi-N as a whole (consisting of Fi-NE and Fi-NW) is a part of the same molecular structure, as is evinced in the channel maps for $^{12}$CO and $^{13}$CO (see Figure~\ref{fig:g_channel}).  Interestingly the $^{12}$CO emission demonstrates vividly that the redshifted Fi-S is indeed a longer filament extending to the northeast, together in the $^{13}$CO a comparatively less prominent feature can also be seen.  The above offers a clear showcase of two filaments moving against each other and has formed a hub, as expected in Stage II or III of the hub-filament paradigm \citep{kum20}.  Moreover the interaction between Fi-NW and Fi-S could be the most plausible mechanism in reddening the velocity in Fi-NW (for details see Appendix~\ref{sec:channel}).  Another possibility could be that Fi-NW is tilted farther away from our line-of-sight compared to that of Fi-NE.  Nonetheless, this inter/intra-filamentary interaction is the initial dominant factor that had led to the formation of the intense and dense hub at the junction of filaments, until when gravity and turbulence kick in at a later stage.

\section{Discussions} \label{sec:discuss}

For more than a decade it has been shown that hubs are the preferred locations for the formation of most massive stars that can form within a filamentary network \citep{mye09, baug18, montillaud19b, kum22}.  However the formation of such massive stars from molecular gas distributed in filamentary structures involves a dynamic range of activities that are still not fully understood or observationally demonstrated.  Here we have identified massive young stars that spatially correlate with the HFS and therefore we interlink them with the dynamics of molecular gas within filaments and investigate their follow-up evolutionary activities.

\subsection{Young stellar Objects}

\begin{figure*}
\centering
        \includegraphics[width=\textwidth]{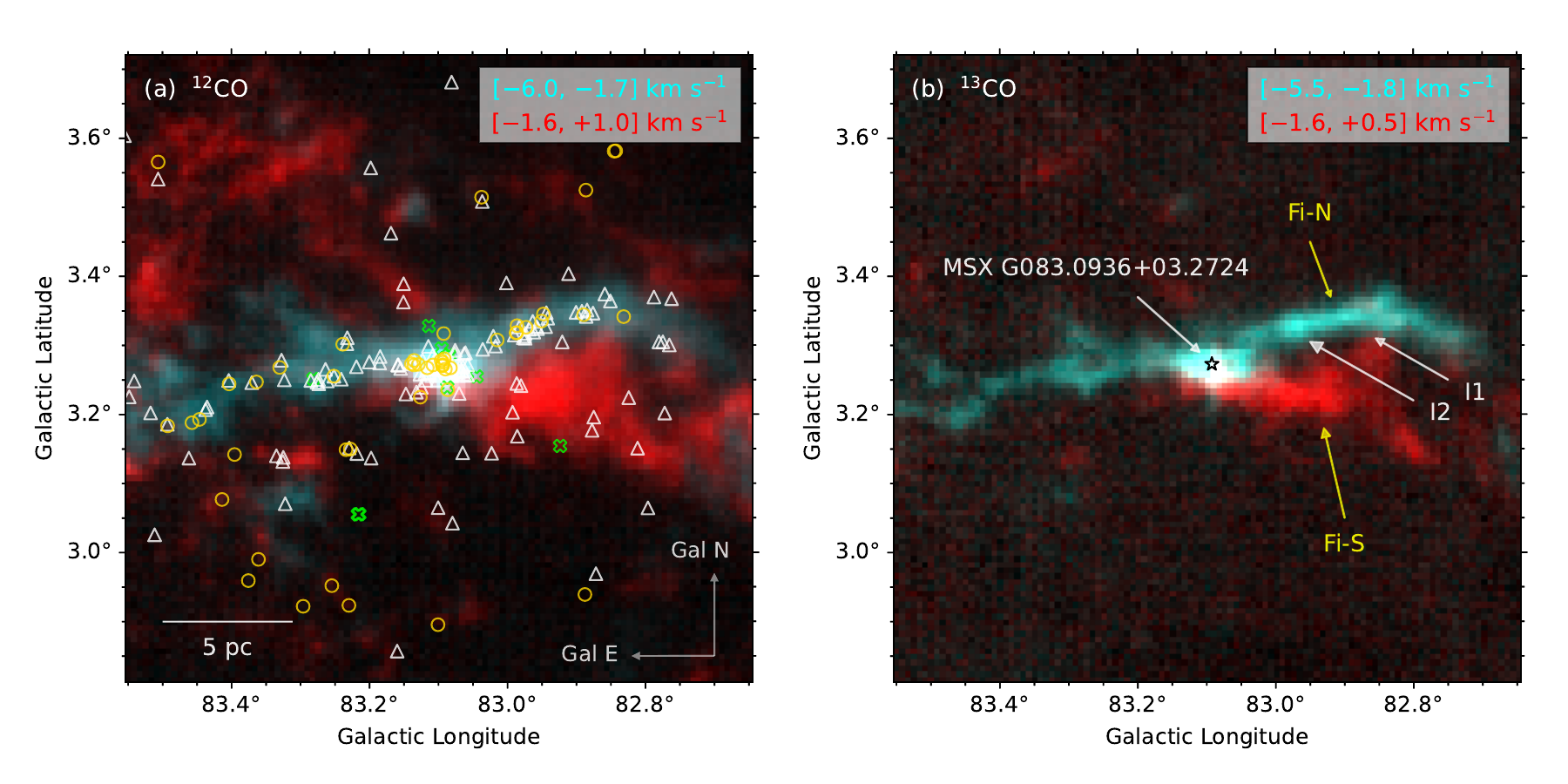}
  \caption{The two color integrated intensity CO images, revealing merging of filaments into a common junction as the hub.  In (a) the young stellar objects (Class~I: yellow circles; Class~II: white triangles; transition-disk: green crosses) from \citet{pan22} are overlaid on the $^{12}$CO distribution.  A scale bar of 5~pc is shown considering a heliocentric distance of 1.52~kpc.  In (b) the skeletons of the two filaments (Fi-N and Fi-S) are more clearly depicted by the $^{13}$CO distribution.  I1 and I2 indicate the subsidiary locations along which Fi-N (western part) and Fi-S are interacting (see Appendix~\ref{sec:channel}).  At the hub, a massive outflow source MSX\,G083.0936$+$03.2724 is marked by a black asterisk symbol. }
  \label{fig:g_rgb}
\end{figure*}

The two color integrated intensity maps presented in Figure~\ref{fig:g_rgb}(a) for $^{12}$CO and in Figure~\ref{fig:g_rgb}(b) for $^{13}$CO demonstrate the elongated filamentary molecular structures for separate velocity ranges.  The young stellar objects (Class~I, Class~II, and transition-disk) toward this region were previously identified \citep{pan22} using the infrared color excess and are overplotted here in Figure~\ref{fig:g_rgb}(a).  The young stars correlate well spatially with Fi-N, with a high concentration of Class~I and Class~II sources populate the hub, indicating enhanced ongoing star formation.  These results are consistent with the higher intensity at the hub, increased by $\sim 5$ times (considering median value for $^{13}$CO integrated intensity) compared to the filaments, caused because of the merging of filaments.  
Whereas Fi-N displays a strongly correlated arrangement of the proto-stellar objects along the entire filament (including Fi-NE and Fi-NW), Fi-S considerably lacks those, possibly due to the fact that Fi-S comprises lower column density (median $\sim 1.2 \times 10^{21}$~cm$^{-2}$ from $^{13}$CO) compared to Fi-NE ($\sim 1.5 \times 10^{21}$~cm$^{-2}$) or Fi-NW ($\sim 2.1 \times 10^{21}$~cm$^{-2}$).  

Moreover, at the hub is located a massive \citep[7.7~M$_{\sun}$,][]{maud15b} outflow (MSX6C\,G083.0936$+$03.2724) with a similar velocity ($V_{\rm lsr} = -3.0$~km~s$^{-1}$).  This outflow source spatially coincides with a Class~I object (2MASS\,20313550$+$4505465) reported in \citet{pan22}.  
Along the hub toward Fi-NW at certain locations (indicated as I1 and I2 in Figure~\ref{fig:g_rgb}(b)), we see density enhancements corresponding with the number of Class~I sources.  
These results are suggestive of an active star formation scenario in G08, caused by the coalescence of molecular gas, thereby channeling material through the filaments and accumulating at the hub.

\subsection{Evolutionary Scenarios}

The filaments Fi-N and Fi-S have lengths of 20.7~pc, and 10.4~pc, respectively, with an overlapping area (hub) of radius $\sim 0.9$~pc.  The width of the filaments varies between 0.7~pc and 2.2~pc, estimated with the $^{13}$CO integrated intensity for corresponding velocity ranges at the $3\sigma$ level.  
Several studies of nearby ($<1$~kpc) high-mass star forming regions have recognized similar types of cluster-forming gas structures and hub morphology, e.g., the central integral shaped filament (ISF) in Orion, which is considered as a hub \citep[radius $\sim 0.8$~pc,][]{hac17} or the Mon\,R2 hub \citep[radius $\sim 0.8$~pc,][]{kum22}, to which G08 reported here has a similar hub size (radius $\sim 0.9$~pc).  Moreover, in the Orion ISF there is evidence of the existence of a global velocity gradient ($\lesssim 1.0$~km~s$^{-1}$~pc$^{-1}$) continuing along the entire filament \citep{hac17}.  
There are increasing observational evidence that these longitudinal velocity gradients in filaments might be involved in driving accretion inflows that feed the central clumps and eventually lead to massive star formation \citep{kir13, per14, hac17, williamsgm18}, as witnessed here in G08.

Near the hub of G08, in addition to the Class~I grouping signifying ongoing star formation, there are also about half a dozen of ionizing stars (early-B) with typical dynamical timescale of $\lesssim 0.2$~Myr \citep{pan22}, based on radio continuum measurements.  
The hub of G08 produces moderately massive stars that are fairly comparable to Mon\,R2 \citep{kum22}, which also contains a cluster of B-type stars but are significantly less massive than the Orion Trapezium Cluster \citep{berne14}.  
In that connection, we infer that the formation of high-mass stars is induced by the interaction of molecular gas at the hub, infalling from the large-scale ($\sim10$~pc) filaments with converging motions.  The velocity gradient along Fi-N is of the order of $\sim 0.084$~km~s$^{-1}$~pc$^{-1}$, which directly reflects inflow of material into the central junction.  Though a massive outflow is located at the hub, we further searched for fresh outflow wings, but given the moderate resolution of the data combined with a relatively larger distance to the region, we did not detect any.  

We found a conspicuous rise both in the excitation temperature (median $\sim 20.4$~K from $^{12}$CO) and the molecular gas column density (median $\sim 2.0 \times 10^{22}$~cm$^{-2}$ from $^{13}$CO) predominantly in and around the hub (see Figure~\ref{fig:g_texmass}(a) and Figure~\ref{fig:g_moment012}(d)).  The measured column density in G08 is well comparable with the hub region in Mon\,R2 \citep[$N_{\mathrm{H_2}} \sim 2$--$3 \times 10^{22}$~cm$^{-2}$, see][Fig. 3]{kum22}, whereas much lower than the Orion-KL \citep[$N_{\mathrm{H_2}} \sim 5 \times 10^{23}$~cm$^{-2}$ from $^{13}$CO,][]{berne14} region, justifiable with the type of stars that are produced in each region.  Hubs that are formed at the junctions of filaments \citep{mye09} essentially provide resources for enhanced star formation due to their density amplification properties \citep{kum20}.  This density amplification in the hub can produce strong gravitational potential difference between the hub and the connected filaments, that in turn is responsible to drive longitudinal flows within the filaments directed inward to the hub \citep{per14, williamsgm18}.  
So the observed HFS morphology in G08 is well consistent with the Stage II or III of the `filaments to clusters' model described by \citet{kum20}, where the two elongated filaments have merged, definitely formed a hub at the intersection and thereafter enhanced the star formation activity therein. 

It is still a subject of discussion about what initiates the filaments to interact at the first instance, however an abundance of anisotropic agents could be liable \citep{hac22}, and the most likely mechanism could be the filaments driven by flow \citep{kum20}.  
Supplementary analyses are also required to compare the effects of gravity, turbulence, and magnetic fields and diagnose the dominating factors in driving the flows and forming the hub in G08.  So this study presents direct observational evidence on filament merging and hub formation and the region has potential to offer ample future avenues for further investigation and enhance our understanding on the dynamical evolution of HFSs.  
Here we reckon that, G08 is in the process of forming massive stars and clusters, with all the favorable circumstances being a reliable candidate for the Stage II or III of the `filaments to clusters' \citep{kum20} paradigm.  With the aid of high resolution interferometric (millimeter and sub-millimeter) and polarimetric (dust continuum and line emission) data, investigation of the central 2~pc hub area would probe deeper insight on the accretion and magnetic activities at such scales.  

\section{Summary and Conclusions} \label{sec:summary}

Our morpho-kinematic study of the filamentary structures in G08 using CO isotopologues reveals the conformity of a HFS that sequentially inducted to the formation of massive young stars and clusters.  
The region harbors two elongated (length $\sim$ 10.4--20.7~pc) filaments with different velocity components ($\sim -3.98$~km~s$^{-1}$ and $\sim -0.83$~km~s$^{-1}$) interacting via a common hub, where we found large velocity spreads (upto $\sim 5.0$~km~s$^{-1}$) with clear bridging features in both $^{12}$CO and $^{13}$CO.  
The significant enhancement in the velocity dispersion at the hub indicates a turbulent gas motion caused by merging of filaments.  
The detection of large-scale velocity gradient ($\sim 0.084$~km~s$^{-1}$~pc$^{-1}$) along the filaments axis indicate a global motion of molecular gas directed inward to the hub and consequently feeding the central clumps.  
The observed molecular gas column density and corresponding star formation in terms of stellar mass in G08 are well comparable with Mon\,R2 and are reasonably much lower than Orion ISF.  
Together, the clusterings of Class I sources, higher excitation temperature (median $\sim 20.4$~K) and higher H$_{2}$ column density (median $\sim 2.0 \times 10^{22}$~cm$^{-2}$ for $^{13}$CO) imply a HFS morphology in G08, where the star formation is similar to Stage II or III of the `filaments to clusters' model described by \citet{kum20}.   
These observational findings adhere convincing evidence on the merging of two giant filaments, seeding the formation of a dense hub and subsequently leading to massive star formation.

\begin{acknowledgments}

We are indebted to the referee for constructive comments and suggestions that have drastically improved the scientific presentation and content of the paper.  
A.P. acknowledges the support provided by the S. N. Bose National Centre for Basic Sciences, funded by the Department of Science and Technology, India.  Y.S. acknowledges support by the Youth Innovation Promotion Association, CAS (Y2022085), and Light of West China Program, CAS (2022-XBJCTD-003).  
This research has made use of the data from the MWISP survey conducted with the PMO-13.7m telescope.  We are grateful to all the members of the MWISP working group, particularly the staff members at PMO, for their long-term support.  MWISP was sponsored by National Key R\&D Program of China with grant 2017YFA0402701 and CAS Key Research Program of Frontier Sciences with grant QYZDJ-SSW-SLH047.  

\end{acknowledgments}

\vspace{5mm}
\facilities{PMO}

\software{GILDAS \citep{gil13}, 
        astropy \citep{astropy13},  
        radio-astro-tools \citep{gin15}
          }

\appendix

\section{Channel Maps and Position-Velocity Diagrams} \label{sec:channel}

\begin{figure*}
\centering
        \includegraphics[width=1.0\textwidth]{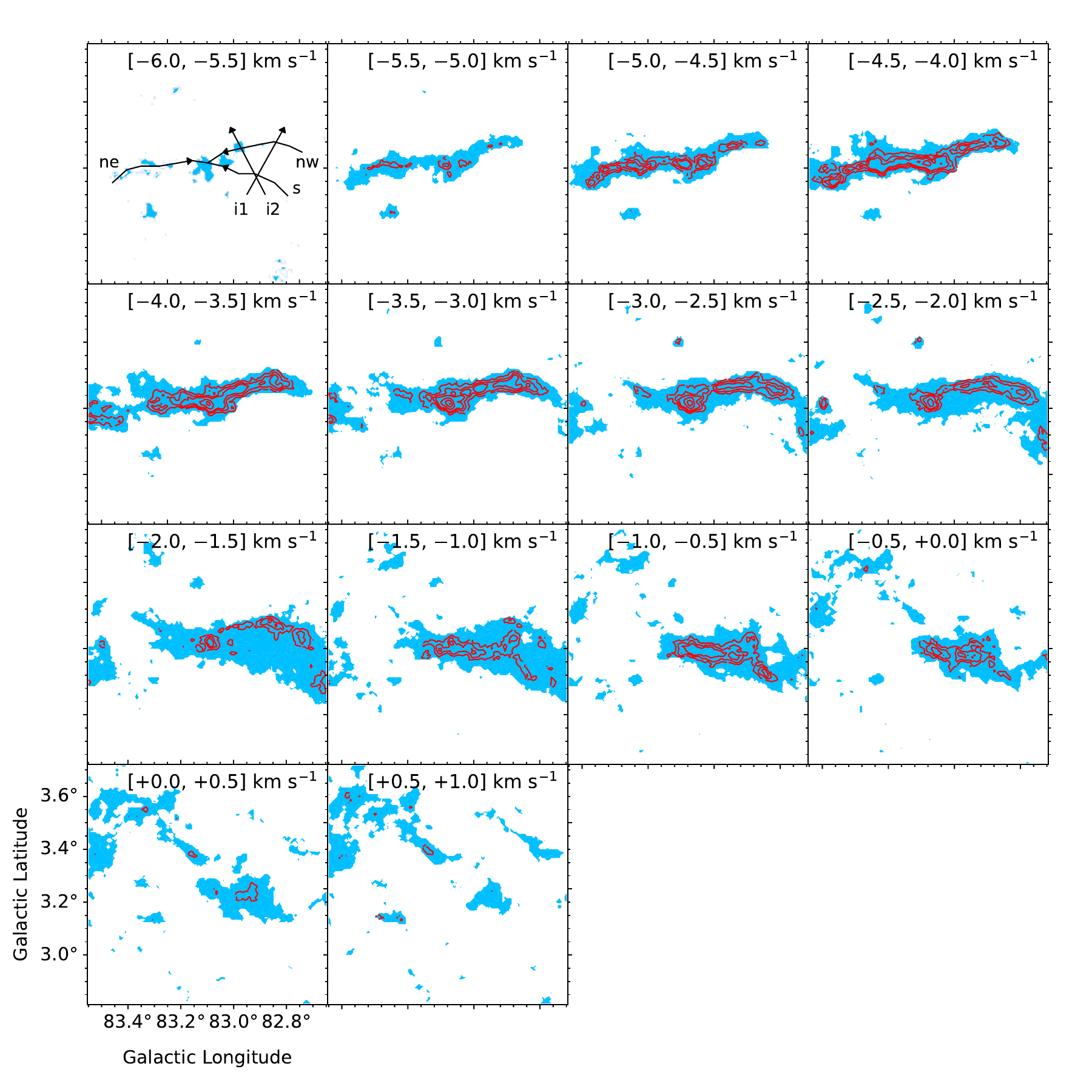}
  \caption{$^{12}$CO channel maps (background) with overlaid of $^{13}$CO contours (red) at the levels of 0.5, 1.0, 1.5, 2.0, and 2.5~K~km~s$^{-1}$ integrated for the mentioned velocity ranges toward G08.  In the first panel, the three black curves (labeled as ne, nw, and s) along the filaments axis and the two black lines (i1 and i2) along the additional interacting zones, each of width $2\farcm5$ (5 pixels) with arrows, indicate the direction along which position-velocity diagrams are generated.  }
  \label{fig:g_channel}
\end{figure*}

The $^{12}$CO and $^{13}$CO channel maps toward the region G08 are presented in Figure~\ref{fig:g_channel}.  We have further investigated the gas motions along filaments axis and at other interaction zones found in-between Fi-NW and Fi-S.  For this, we have chosen total five paths, three along filaments axis (referred as ne, nw, and s) all directed inward to the hub, and two along additional interacting zones (referred as i1 and i1), each for a width of $2\farcm5$ (5 pixels), as depicted in Figure~\ref{fig:g_channel}.  

\begin{figure*}
\centering
        \includegraphics[width=\textwidth]{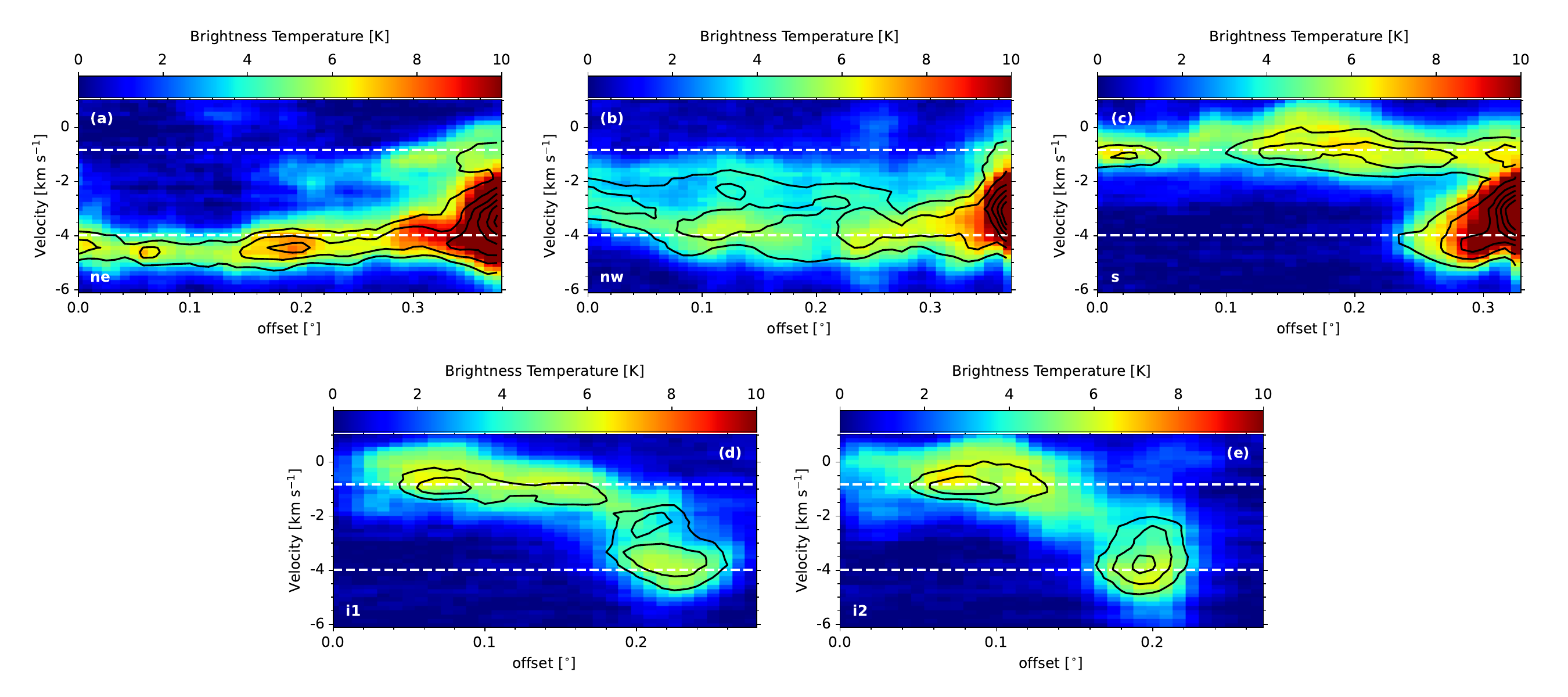}
  \caption{$^{13}$CO position-velocity contours (black) are overlaid on the $^{12}$CO position-velocity diagrams along (a) ne, (b) nw, (c) s, (d) i1, and (e) i2, as mentioned in Figure~\ref{fig:g_channel}.  The contour levels are at 1, 2, 3, 4, 5, and 6~K, respectively.  The white dashed lines mark velocities of $-3.98$~km~s$^{-1}$ and $-0.83$~km~s$^{-1}$, respectively.  }
  \label{fig:g_pv}
\end{figure*}

The corresponding position-velocity diagrams along the five defined paths (as indicated in Figure~\ref{fig:g_channel}) for both $^{12}$CO and $^{13}$CO are shown in Figure~\ref{fig:g_pv}.  Toward Fi-NE (Figure~\ref{fig:g_pv}(a)), the velocity is bluer and gradually shifts between $\sim -4.31$~km~s$^{-1}$ (east) to $\sim -3.35$~km~s$^{-1}$ (hub), consistent with reported by  \citet{pan22}.  Overall the velocity spread is also relatively low  ($\sim 1.0$~km~s$^{-1}$, at $3\sigma$ level of $^{13}$CO emission), except around the hub, where the spread is significantly higher ($\sim 5.0$~km~s$^{-1}$), understandably as in the case of a hub in a traffic network, bridging multiple velocity components from different filaments.  
In contrast, toward Fi-NW (Figure~\ref{fig:g_pv}(b)), the velocity increases slightly, varying in the range $\sim -3.35$~km~s$^{-1}$ (hub) to $\sim -2.41$~km~s$^{-1}$ (west), with a noticeable spread of ($\sim 3.0$~km~s$^{-1}$) along the filament, especially at the intermediate zones, as if the gas within are highly disturbed by external influences.  On that account, moving from the eastern to the western part of Fi-N, keeping the hub roughly at the center, we can see velocity gradients indicating a global gas motion, whereas in Fi-S (Figure~\ref{fig:g_pv}(c)), the velocity peak is different ($\sim -0.83$~km~s$^{-1}$) with a single component and has a moderate spread.  We therefore infer that there are actually two major velocity components, $\sim -3.98$~km~s$^{-1}$ (along Fi-N) and $\sim -0.83$~km~s$^{-1}$ (along Fi-S), while the hub and Fi-NW (partially) accommodate a combination of those.  
Along I1 (Figure~\ref{fig:g_pv}(d)) and I2 (Figure~\ref{fig:g_pv}(e)), the molecular gas in Fi-NW and Fi-S is found to be clearly inter-connected, as these bridge the two main velocity components, therefore providing compelling evidence of physically association.

\section{Derivation of H$_{2}$ Column Density and H$_{2}$ Mass} \label{sec:column}

Toward G08, we have computed the H$_{2}$ column density and the H$_{2}$ mass for both $^{13}$CO and C$^{18}$O for the velocity range of [$-6.0$, $+1.0$]~km~s$^{-1}$, using a similar methodology outlined in \citet{sun20} and the procedure is briefed below.  Initially the excitation temperature is derived from the peak brightness temperature of the optically thick $^{12}$CO line and the resultant map is displayed in Figure~\ref{fig:g_texmass}(a).  Following which, the optical depths for the corresponding isotopologues ($^{13}$CO and C$^{18}$O) are derived (maps are not shown here) by assuming that they have same excitation temperature.  The obtained values are further used to compute the H$_{2}$ column density by assuming local thermodynamic equilibrium (LTE) approximation and abundance ratios of H$_{2}$ to $^{13}$CO and C$^{18}$O.  The H$_{2}$ mass (for both $^{13}$CO and C$^{18}$O) are then obtained by integrating the column density maps and assuming the LTE method.  The H$_{2}$ mass distribution for $^{13}$CO is shown in Figure~\ref{fig:g_texmass}(b).  For these processes, equations (1)--(5) from \citet{sun20} are incorporated.

\section{Ionized Emission and Dense Clumps} \label{sec:clump}

\begin{figure*}
\centering
        \includegraphics[width=1.0\textwidth]{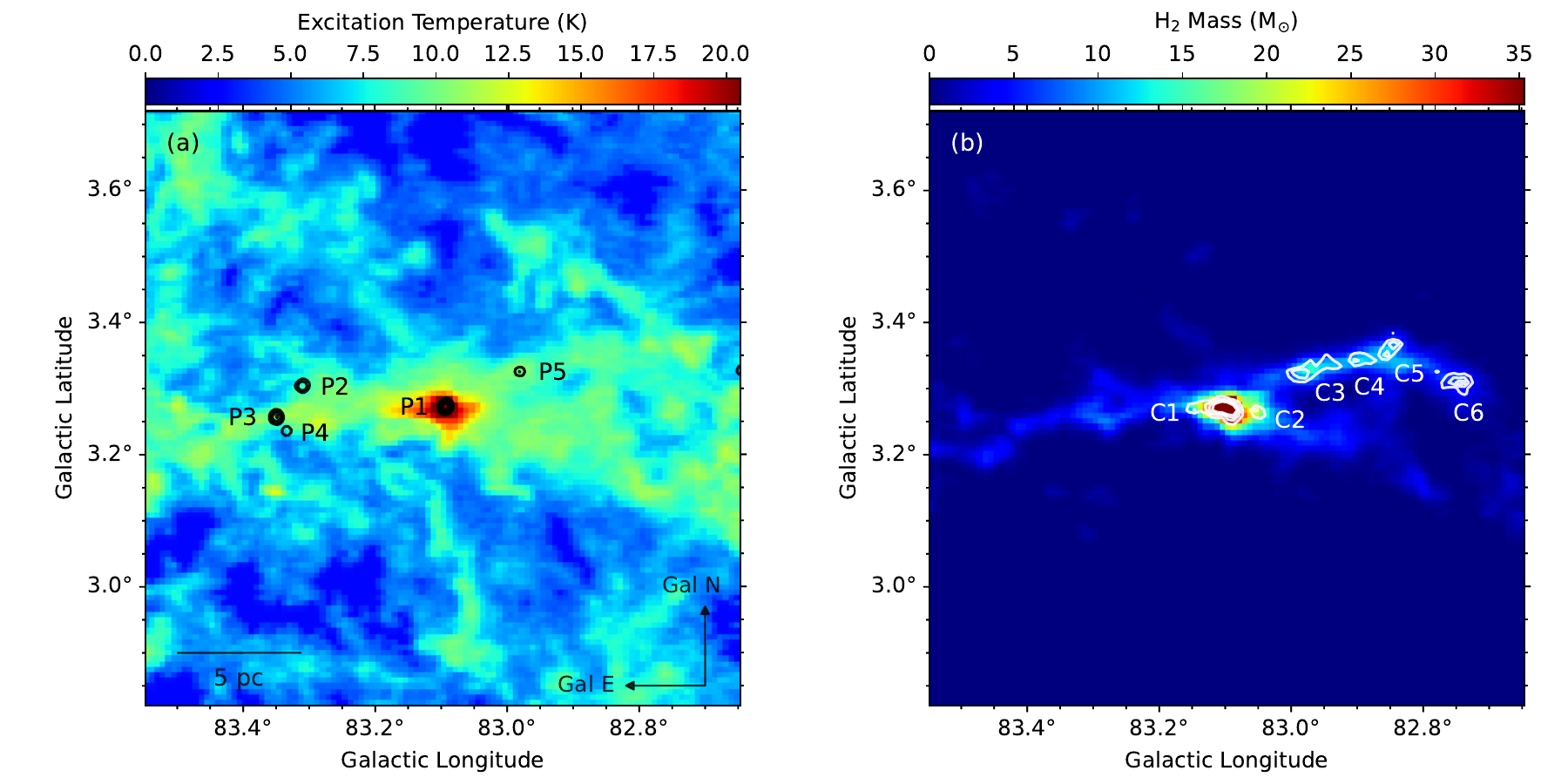}
  \caption{(a) The excitation temperature map derived from $^{12}$CO.  The radio continuum emission from the NVSS 1.4~GHz is depicted in black contours at levels of 0.003, 0.006, 0.012, and 0.020~Jy~beam$^{-1}$.  A total of five emission peaks with integrated flux densities above the $3\sigma$ level are detected and are marked as P1 to P5.  (b) The H$_{2}$ mass distribution derived from $^{13}$CO in the velocity range [$-6.0$, $+1.0$]~km~s$^{-1}$.  The C$^{18}$O intensity map integrated for the same velocity range, as overlaid as white contours at levels of 0.199, 0.398, 0.597, and 0.796~K~km~s$^{-1}$.  Six massive clumps labeled as C1 to C6 are identified from the C$^{18}$O data.  }
  \label{fig:g_texmass}
\end{figure*}

We have traced five ionized emission peaks (above $3\sigma$) using the NVSS 1.4~GHz \citep{con98} radio continuum data, shown as contours in Figure~\ref{fig:g_texmass}(a) and are marked as P1 to P5, with the flux density varying in the range $\sim 7$--$33$~mJy.  

We have looked for C$^{18}$O emission and identified six massive clumps located at the hub and toward Fi-NW, labeled as C1 to C6 in Figure~\ref{fig:g_texmass}(b), with their parameters summarized in  Table~\ref{tab:g08_clump}.  
The clumps C1 and C3 coincide, respectively, with the ionization peaks P1 and P5.  Some of the clumps show signs of further fragmentation into smaller cores.  Interestingly the clumps C1, C3, and C5 also correlate with the position of the hub, and with the interacting zones I1 and I2.

\begin{deluxetable*}{ccccc}
\tablecaption{Molecular properties of the clumps obtained from C$^{18}$O}
\label{tab:g08_clump}
\tablehead{
\colhead{Clump} & \colhead{Glon.} & \colhead{Glat.} & \colhead{$\rm H_{2}$ Mass} & \colhead{Deconvolved Radius}  \\
\colhead{ID} & \colhead{(deg)} & \colhead{(deg)} & \colhead{($M_{\sun}$)} & \colhead{(pc)} 
           }
\startdata
C1  &  083.1026  &  +03.2693  &  $305.2 \pm 17.5$   &  0.74  \\
C2  &  083.0503  &  +03.2650  &  $26.6 \pm 5.0$   &  0.34  \\
C3  &  082.9829  &  +03.3212  &  $48.3 \pm 6.9$   &  0.61  \\
C4  &  082.8950  &  +03.3422  &  $25.3 \pm 5.0$   &  0.42  \\
C5  &  082.8513  &  +03.3595  &  $30.7 \pm 5.5$   &  0.48  \\
C6  &  082.7476  &  +03.3099  &  $44.0 \pm 6.6$   &  0.56  \\
\hline 
\enddata
\end{deluxetable*}

\bibliography{ms_g08}{}
\bibliographystyle{aasjournal}



\end{document}